\documentclass[12pt]{article}
\usepackage{epsfig,amsfonts,amssymb}
\usepackage{hyperref}

\usepackage{cite}
\usepackage{cite}

\usepackage{comment}

\def\ZZZ{{\hbox{ Z\kern-1.6mm Z}}}
\def\RRR{{\hbox{ R\kern-2.4mm R}}}
\def\CCC{{\hbox{ C\kern-2.0mm C}}}
\def\zzz{{\hbox{z\kern-1mm z}}}

\newcommand{\qeq}{{\hbox{=\kern-2.3mm ? \kern.5mm }}}
\renewcommand{\qeq}{=}

\newcommand{\vp}{\varphi}

\newcommand{\AAA}{{\cal A}}

\newcommand{\wt}{\widetilde}
\newcommand{\wh}{\widehat}

\newcommand{\NN}{{\cal N}}

\newcommand{\be}{\begin{equation}}
\newcommand{\ee}{\end{equation}}
\newcommand{\ben}{\begin{eqnarray}\displaystyle}
\newcommand{\een}{\end{eqnarray}}

\newcommand{\refb}[1]{(\ref{#1})}
\newcommand{\p}{\partial}
\newcommand{\sectiono}[1]{\section{#1}\setcounter{equation}{0}}

\def\one{{\hbox{ 1\kern-.8mm l}}}
\def\zero{{\hbox{ 0\kern-1.5mm 0}}}

\newcommand{\bea}[1]{\begin{eqnarray}\label{#1} }
\newcommand{\eea}{\end{eqnarray}}

\newcommand{\bN}{{\bf N}}

\newcommand{\eqref}{\refb}

\usepackage{bm}
\usepackage[table]{xcolor}

\def\bN{{\bf N}}

\def\figpicardthree{

\def\JPicScale{0.8}
\ifx\JPicScale\undefined\def\JPicScale{1}\fi
\unitlength \JPicScale mm
\begin{picture}(130,80)(0,0)
\linethickness{0.1mm}
\put(70,0){\line(0,1){80}}
\linethickness{0.1mm}
\put(10,40){\line(1,0){120}}
\linethickness{0.3mm}
\linethickness{0.3mm}
\linethickness{0.3mm}
\linethickness{0.3mm}
\linethickness{0.3mm}
\linethickness{0.3mm}
\linethickness{0.5mm}
\qbezier(70,40)(80.38,40)(90,40)
\qbezier(90,40)(99.62,40)(110,40)
\qbezier(110,40)(120.44,40)(125.25,40)
\qbezier(125.25,40)(130.06,40)(130,40)
\linethickness{0.5mm}
\qbezier(70,40)(70,37.4)(70,35.59)
\qbezier(70,35.59)(70,33.79)(70,32.5)
\qbezier(70,32.5)(70.02,31.23)(68.81,27.62)
\qbezier(68.81,27.62)(67.61,24.02)(65,17.5)
\qbezier(65,17.5)(62.41,10.99)(60,6.78)
\qbezier(60,6.78)(57.59,2.57)(55,0)

\put(95,40){\makebox(0,0)[cc]{$\times$}}

\put(95,45){\makebox(0,0)[cc]{$\psi^0=1/a$}}

\put(95,65){\makebox(0,0)[cc]{complex $\psi^0$-plane}}

\put(60,45){\makebox(0,0)[cc]{$\psi^0=0$}}

\put(80,39.8){\makebox(0,0)[cc]{$\Rightarrow$}}

\put(120,39.8){\makebox(0,0)[cc]{$\Rightarrow$}}

\put(70,40){\makebox(0,0)[cc]{$\times$}}

\put(70,35){\makebox(0,0)[cc]{$\Uparrow$}}

\end{picture}

}

\begin{document}

\baselineskip 24pt

\begin{center}

{\Large \bf Normalization of D-instanton Amplitudes}

%{\Large \bf Classical Limit of Soft Theorems in Arbitrary Dimensions}

\end{center}

\vskip .6cm
\medskip

\vspace*{4.0ex}

\baselineskip=18pt

\centerline{\large \rm Ashoke Sen}

\vspace*{4.0ex}

\centerline{\large \it Harish-Chandra Research Institute, HBNI}
\centerline{\large \it  Chhatnag Road, Jhusi,
Allahabad 211019, India}

%\centerline{\large \it ~$^c$Homi Bhabha National Institute}
%\centerline{\large \it Training School Complex, Anushakti Nagar,
%    Mumbai 400085, India}

\vspace*{1.0ex}
\centerline{\small E-mail:  sen@hri.res.in}

\vspace*{5.0ex}

\centerline{\bf Abstract} \bigskip

D-instanton amplitudes suffer from various infrared divergences associated with tachyonic
or massless open string modes, leading to ambiguous contribution to string amplitudes.
It has been shown previously 
that string field theory can resolve these ambiguities and lead
to unambiguous expressions for D-instanton contributions to string amplitudes, except for an
overall normalization constant that remains undetermined. In this paper we show
that string field theory, together with the world-sheet description of the amplitudes, 
can also fix this normalization constant. We apply our analysis to the
special case of two dimensional string theory, obtaining results in 
agreement with the matrix model
results obtained by Balthazar, Rodriguez and Yin.

\vfill \eject

\tableofcontents

\sectiono{Introduction} \label{s1}

D-instantons give a class of non-perturbative contributions to string amplitudes.
One  characteristic  of these contributions is the presence of an 
overall multiplicative factor $e^{-C/g_s}$
where $g_s$ is the closed string coupling and $C$ is a 
constant. Besides this factor, the amplitudes admit usual perturbation 
expansion in powers of $g_s$.  The contribution to an amplitude at any given order
in $g_s$ can be computed 
using the standard world-sheet approach by including Riemann surfaces with boundaries
ending on the D-instanton, but at each order 
one encounters certain infra-red divergences\cite{9407031,9701093,1907.07688} 
that render the 
amplitudes ambiguous. At any given order,
these ambiguities can be 
encoded in a set of undetermined constants. 
String field theory\cite{wittensft,9206084,9705241,1703.06410,1905.06785,okawa} provides an 
unambiguous procedure for determining these constants, by identifying the
physical origin of these infrared divergences and rectifying them based on this 
understanding\cite{1908.02782,2002.04043,2003.12076,2012.11624}. 
So far this procedure 
has been applied to two dimensional 
string theory, for which there is a dual matrix model description that can be used to check
the results.

However previous analysis left one constant undetermined -- namely the overall
normalization of the D-instanton amplitude. Formally this is given by the exponential of the
annulus amplitude, with D-instanton boundary condition at the two boundaries and no
other vertex operator insertion. However the annulus amplitude is divergent due to the
presence of massless and tachyonic open string modes on the D-instanton. 
In conventional string perturbation theory, 
such diagrams are part of bubble diagrams and
drop out in the computation of physical amplitudes. However for D-instanton amplitudes
the situation is somewhat different since the D-instanton contribution to the amplitude
has to be first added to the perturbative amplitude and then the sum needs to be divided
by the sum of perturbative and D-instanton contribution to 
bubble diagrams. Therefore the overall normalization is physically relevant,
and one expects that it should be possible to compute this in string theory. Since string
field theory is capable of making sense of infrared divergences in the amplitudes, the
natural expectation would be that string field theory should be able to give an
unambiguous result for the normalization constant.
However, when one tries to compute this using string field theory, which in this
case is a theory of open and closed strings, one finds that there is no internal consistency
requirement within string field theory that can be used to fix this normalization, since this
can be changed by adding a field independent constant to the string field theory action
that does not violate any constraint coming from the requirement of gauge invariance.

To overcome this problem, we shall take the viewpoint that the world-sheet approach already
fixes the normalization as the exponential of the annulus partition function, and the job of string
field theory is to simply give physical interpretation of the divergences of the amplitude and
render them finite based on this interpretation.
We show that the world-sheet result may be regarded as the gauge fixed version of a path 
integral in string theory with a specific normalization, 
and the divergences that we encounter arise due to breakdown
of the gauge choice. However the `gauge invariant' form of the path integral, expressed
as an integral over the full classical string field divided by the volume of the gauge group,
yields unambiguous result.
We apply this procedure
to the case of 
two dimensional string theory, and find that the normalization of the one instanton amplitude
determined this way agrees with the results of the matrix model computed in \cite{1907.07688}
following the general formalism developed in \cite{9111035}.

The rest of the paper is organized as follows. In \S\ref{s2} we 
express the exponential of the annulus partition function as a path integral over string fields
in the Siegel gauge. At this stage the path integral remains singular due to the existence of
zero modes, reflecting the singularity of
the annulus partition function. 
In \S\ref{s3} we 
trace these singularities to the breakdown of the Siegel gauge, and show that we can get
finite result for the path integral by rewriting it in a `gauge invariant' form.
In \S\ref{s3.5} we calculate the multiplier factor -- the multiple of the steepest descent contour
of the D-instanton that forms part of the actual integration contour of the full string theory, and
show that after multiplying the exponential of the annulus partition function, computed in
\S\ref{s3}, by this factor, we get agreement 
with
the matrix model result. In \S\ref{s4} we discuss possible application of our analysis to other
systems. In appendix \ref{sa} we show how the central result used in our analysis -- the
equivalence of the gauge invariant version of the path integral over string fields and the
Siegel gauge fixed version of the same path integral, can be proved directly using the
standard Faddeev-Popov approach instead of the abstract results of the Batalin-Vilkovisky
(BV) formalism\cite{bv1,bv2,henn,bocc1,bocc2,thorn,9205088}.

\sectiono{The normalization constant from the Siegel gauge path integral} \label{s2}

\begin{figure}
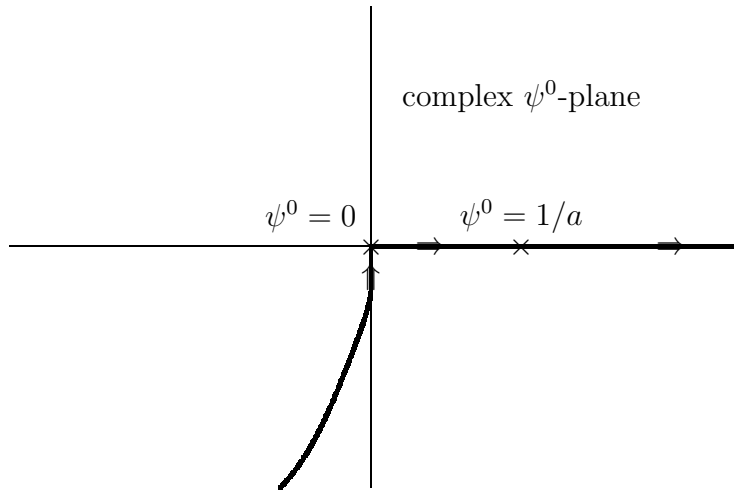

\begin{center}
\figpicardthree
\end{center}

\vskip -.2in

\caption{The integration contour
in the complex $\psi^0$ plane.
\label{figpicardthree}
}
\end{figure}

Our goal is to compute the normalization constant $\bN$ appearing in the D-instanton
amplitudes. In the world-sheet description it is given by:
\be \label{e1}
\bN = \zeta\, \exp[\AAA]\, ,
\ee
where  $\AAA$ is the exponential
of the annulus partition function:
\be \label{e2n}
\AAA = \int_0^\infty {dt\over 2t} \, Tr(e^{-2\pi t L_0})\, .
\ee
Here $Tr$ denotes trace over all states of the open string projected into the Siegel gauge by
the projection operator $b_0c_0$,
weighted by $(-1)^F$ where $-(-1)^F$ denotes the grassmann
parity of the vertex operator corresponding to the state.  The extra minus sign multiplying
$(-1)^F$ is a reflection of the fact that bosonic (fermionic) open string modes correspond
to grassmann odd (even) states in the world-sheet theory.
$\zeta$ in \refb{e1} is the multiplier factor that depends on how the
steepest descent contour associated with the D-instanton fits inside the actual
integration contour\cite{1206.6272,1511.05977,1802.10441}. 
In particular $\exp[\AAA]$ represents the one loop contribution to the
path integral from the full steepest descent contour passing through the instanton
solution and
$\zeta$ reflects the multiple of the steepest descent contour that forms part of the
actual integration contour.
We shall see for example that in the two dimensional string theory, $\zeta=1/2$
up to a sign.\footnote{In string field theory, \refb{e1} 
may be justified by demanding
that at the tachyon vacuum\cite{9911116} 
$\bN$ must be 1 so that we get the usual
perturbative closed string amplitudes. Since the boundary state vanishes at the tachyon vacuum\cite{0810.1737}, 
we have $\AAA=0$ and therefore $e^\AAA=1$. Furthermore it will be seen in
\S\ref{s3.5}, Fig.~\ref{figpicardthree} that $\zeta=1$ at the perturbative vacuum. Therefore
\refb{e1} should not have any additional factor.}

The constant $\bN$ given in \refb{e1}
is the overall multiplicative factor that appears in the instanton
induced effective action of the closed string fields\cite{ 2012.00041}. 
This is related to the normalization constant $\NN$ introduced
in \cite{1907.07688,1912.07170}, appearing 
as a multiplicative factor in the S-matrix, via the relation $\NN=i\,\bN$.
In our analysis, we shall not be careful in fixing the
sign of $\bN$, since this will be fixed at the end using separate considerations.

We can express \refb{e2n} as
\be\label{e3}
\AAA = \int_0^\infty {dt\over 2t} \left[\sum_i e^{-2\pi t h^b_i} -\sum_j e^{-2\pi t h^f_j}\right]\, ,
\ee
where $\{h^b_i\}$ and $\{h^f_j\}$ are the $L_0$ eigenvalues of the grassmann odd 
and the grassmann even
states of the world-sheet CFT. 
If we assume that the total number of bosonic modes equals
the total number of fermionic modes so that the integrand is finite as $t\to 0$, and furthermore
that the $h^b_i$ and $h^f_j$ are positive so that the integrand falls off exponentially as
$t\to\infty$, then the integral \refb{e3} is finite. In this case it gives the result
\be 
\AAA={1\over 2} \ln {\prod_j h^f_j\over \prod_i h^b_i}\, .
\ee
Substituting this into \refb{e1}, we get
\be
\bN = \zeta\, \sqrt{\prod_j h^f_j\over \prod_i h^b_i}\, .
\ee
In the system that we shall analyze, 
$h^f_j$'s come in pairs of equal values so that we can write this as
\be \label{e8pre}
\bN = \zeta\, {\prod'_j h^f_j\over \sqrt{\prod_i h^b_i}}\, ,
\ee
where $\prod'_j$ corresponds to the product running over only one 
member for each pair.

We can express this as the result of integration over the bosonic variables $b_i$ and 
fermionic variables $f_j,\wt f_j$ as follows:
\be\label{e8}
\bN =\zeta\, 
\int \prod_i \, {db_i \over \sqrt{2\pi}} 
\prod_j df_j \, d\wt f_j \, \exp\left[-{1\over 2} \sum_i h^b_i b_i^2
+ {\sum_j}' \, h^f_j \, \wt f_j \, f_j\right] \, .
\ee
Equality of \refb{e1} and \refb{e8} is an identity when all the $h^b_i$'s and $h^f_j$'s are positive,
but we shall take \refb{e8} to be the defining expression for $\bN$ even when this condition 
fails.
In particular,
we shall apply this formalism to D-instanton system for which some of the $L_0$
eigenvalues
vanish and / or are negative. 
A justification for this may be given as follows. 
Instead of studying open strings on a single D-instanton, we can
take a system of two D-instantons separated along the Euclidean 
time direction and analyze the states of the open string stretched between the pair of
D-instantons. In this case $L_0$ will get a non-vanishing contribution from the tension of
the stretched open string and the manipulations carried out above will be well defined for
sufficiently large separation.
We can recover the original system of interest by analytic continuation of this
result to zero separation and 
using the  fact that in this limit the spectrum of open strings with two ends lying on
different D-instantons coincides with the spectrum of open strings with both ends
lying on the same D-instanton. 
Of course \refb{e8} is not
well defined in this limit due to the appearance of zero eigenvalues in the bosonic 
and fermionic
sectors, and so it does not lead to a finite unambiguous result for $\bN$ at this stage. 
However,  we shall see in \S\ref{s3} that it is possible to trace these zero eigenvalues to
singular gauge choice and transform \refb{e8} to 
finite, unambiguous result \refb{egaugeinv} using 
insights from string field theory.

In \refb{e8},
the variables $b_i$ may be interpreted as
the bosonic open string fields on the D-instanton, the variables $f_i,\wt f_i$ may be
interpreted as the 
fermionic open string 
fields on the D-instanton and the argument of the exponential may be
interpreted as the
quadratic part of the action of the open string field theory in the 
Siegel gauge.
To see how this arises, we now review some basic aspects of string field theory.

The off-shell open string field describing the degrees of freedom of a D-instanton 
is taken to be an arbitrary
element $|\Psi\rangle$ of $H$ --
the vector space of states of the open string, including
matter and ghost excitations.  Let $\{|\phi_r\rangle\}$ be the set of basis
states in $H$.  Then we can expand $|\Psi\rangle\in H$ as
\be\label{esftexpansion}
|\Psi\rangle =\sum_r \chi^r |\phi_r\rangle\, .
\ee
$\{\chi^r\}$'s are the degrees of freedom over which the path integral
is to be performed after suitable gauge fixing. 
Even though we have referred to the $\chi_r$'s as fields, they 
are actually zero dimensional fields -- ordinary variables -- since 
on the D-instanton the open strings do not carry any continuous momentum labels.
Therefore it is more appropriate to call them modes.
$\chi^r$ has even (odd) grassmann parity if the ghost number of $\phi_r$ is
odd (even).
The kinetic term of the BV master action of string 
field theory takes the form:
\be\label{esftaction}
S=-{1\over 2} \langle \Psi|Q_B|\Psi\rangle\, , 
\ee
where $Q_B$ is the world-sheet BRST operator. The minus sign in front of the
action is unusual, but has been introduced keeping in mind that we shall be using a 
convention in which the Euclidean path integral is weighted by $e^S$.

In the BV formalism
the open string modes
multiplying states of ghost number $\le 1$ are regarded 
as fields and the modes multiplying
states of ghost number $\ge 2$ are regarded as antifields. 
If we introduce basis states $\{|\vp_r\rangle\}$ in the ghost number
$\le 1$ subspace and $\{|\vp^r\rangle\}$ in the ghost number $\ge 2$ subspace such that
\be \label{eortho}
\langle \vp^r |\vp_s\rangle=\delta^r_s=\langle\vp_s|\vp^r\rangle, \qquad 
\langle\vp^r|\vp^s\rangle
=0, \qquad \langle \vp_r|\vp_s\rangle=0\, ,
\ee
and expand the string field as,
\be \label{edefantifield}
|\Psi\rangle = \sum_r( \psi^r |\vp_r\rangle + \psi_r |\vp^r\rangle)\, ,
\ee
then we call $\psi^r$ a field and $\psi_r$ the conjugate anti-field up to a sign.
The path integral 
is carried out over a Lagrangian submanifold. 
For our analysis it will be sufficient to consider a special class of 
Lagrangian submanifolds in which, 
for each pair $(\psi^r, \psi_r)$, we set either $\psi^r$ to 0 or $\psi_r$ to 0.  
The path integral can be shown to be
formally independent of the choice of the Lagrangian submanifold.
The Siegel gauge corresponds to the choice of the Lagrangian submanifold
in which we impose the condition:
\be \label{esiegelgauge}
b_0|\Psi\rangle=0\, .
\ee
In this gauge the action \refb{esftaction} takes the form:
\be \label{eactiongf}
S_{g.f.}= -{1\over 2} \langle \Psi|c_0 L_0|\Psi\rangle \,.
\ee
If we choose the basis states $\{|\phi^{(n)}_r\rangle\}$ of ghost number $n$ in the Siegel
gauge, satisfying
\be 
b_0|\phi^{(n)}_r\rangle =0,  \qquad \langle\phi^{(2-n)}_r|c_0|\phi^{(n)}_s\rangle =\delta_{rs}
\quad \hbox{for $n\le 1$}\, ,
\ee
then by expanding $|\Psi\rangle$ in this basis and substituting in the action 
\refb{eactiongf}, we recover the exponent in \refb{e8} if we identify the variables
$b_i,f_i$ and $\wt f_i$ as the coefficients of expansion of $|\Psi\rangle$ in this basis.

This shows that \refb{e8} may be given an interpretation as path integral over the
open string fields in the Siegel gauge.
Note however that \refb{e8} comes with a specific normalization 
of the integration measure that
will be important for us. String field theory, by itself, cannot fix the overall normalization
of the measure, since this corresponds to adding a constant to the string field theory action,
and the requirement that the action satisfies the BV master equation 
does not fix this constant.

\sectiono{`Gauge invariant' path integral} \label{s3}

Let us now focus on the specific case of (1,1) D-instanton in two dimensional string theory.
In this case we have\cite{1912.07170}:
\be
\AAA = \int_0^\infty {dt\over 2t} \, \left( e^{2\pi t}-1\right)\, .
\ee
Comparing this with \refb{e3} we see that the contribution from all states with $L_0>0$
cancel between bosonic and fermionic states. 
It follows that in the path integral expression \refb{e8} for $\bN$, we can drop the
integration over all states with $L_0>0$. Therefore we shall introduce a restricted string
field $|\Psi_R\rangle$
given by a linear combination of basis states with $L_0\le 0$. Before gauge fixing,
$|\Psi_R\rangle$ has the following expansion:
\ben \label{epsir}
|\Psi_R\rangle &=& \psi^0 c_1|0\rangle + \psi_0 c_0c_1|0\rangle + 
\psi^1 c_0|0\rangle + \psi^2|0\rangle + \psi_1 \, c_{-1} c_1|0\rangle + \psi_2 \, c_{-1}
c_0 c_1|0\rangle \nonumber \\ && + \, \psi^3 c_1\alpha_{-1}|0\rangle + \psi_3 c_0c_1 \alpha_{-1} |0\rangle\, ,
\een
where $|0\rangle$ is the SL(2,R) invariant vacuum, $c_n$, $b_n$ are the usual ghost oscillators
and
$\alpha_m$ are the oscillators associated with the Euclidean time coordinate $X$, satisfying
$[\alpha_m,\alpha_n]=m\, \delta_{m+n,0}$. 
In the $\alpha'=1$
unit the $X$'s satisfy the operator product expansion\footnote{We shall use the standard
doubling trick in which we regard $\p X$ as an analytic function over the full complex plane, with
the understanding that $\p X(z)$ for $z$ in the lower half plane actually represents 
$-\bar\p X(z)$. \label{fo1}}
\be\label{exxope}
\p X(z) \, \p X(w) = -{1\over 2 (z-w)^2}\, .
\ee
This leads to the following state operator correspondence:
\be \label{exxope1}
c_1\alpha_{-1}|0\rangle = i\, \sqrt 2 \, c(0) \, \p X(0)|0\rangle\, .
\ee
The basis states in which we have expanded the
string field in \refb{epsir} 
are normalized according to \refb{eortho} provided we choose:
\be
\langle 0| c_{-1} c_0 c_1|0\rangle =1\, .
\ee 
In this case the $\{\psi^r\}$'s label
fields and the $\{\psi_r\}$'s label the conjugate anti-fields in the BV formalism. 

In the Siegel gauge, the modes that survive are $\psi^0,\psi_1,\psi^2$ and $\psi^3$. Of these, 
$\psi^0$ and $\psi^3$ are bosonic modes and $\psi_1$ and $\psi^2$ are fermionic
modes. 
Therefore, \refb{e8} may now be written as:
\be\label{esigact}
\bN = \zeta\, \int {d\psi^0\over \sqrt{2\pi}} \int {d\psi^3\over \sqrt{2\pi}} \, 
d\psi_1 \, d\psi^2 \, e^S\, .
\ee
However, as
discussed in \cite{2002.04043,2012.11624}, 
this is a singular gauge choice due to the presence of the zero mode
$\psi_1$.
We avoid this problem by choosing the `gauge' in which $\psi_1=0$ but 
$\psi^1\ne 0$. In this case all the anti-fields are set to zero and we
we integrate over all the field modes $\psi^0,\psi^1,\psi^2,\psi^3$.\footnote{Consequences
of this for tree amplitudes have been discussed in 
\cite{1912.05463,2002.04043,2006.16270}.}
Now
we have three bosonic modes $\psi^0,\psi^1,\psi^3$ and one fermionic mode
$\psi^2$.
The action \refb{esftaction} now takes the form:
\be \label{eresaction}
S = - \left[ -{1\over 2} (\psi^0)^2 - (\psi^1)^2
\right]\, .
\ee
This
shows that we still have a pair of zero modes -- one bosonic zero mode $\psi^3$ and
one fermionic zero mode $\psi^2$ over which we need to integrate. 
Since $\psi^2$ is a grassmann odd variable,
naively the integral would vanish. However,  the mode
$\psi^2$ is the ghost field associated with the string field theory gauge transformation 
generated by $\theta|0\rangle$ for a parameter $\theta$, and the integration over $\psi^2$
can be interpreted as division by $\int d\theta$, with the integral running over the volume of
the gauge group\cite{2002.04043,2012.11624},. 
This allows us to express \refb{esigact} as,
\be \label{egaugeinv}
\bN =\zeta\,  \int {d\psi^0\over \sqrt{2\pi}} \int {d\psi^3\over \sqrt{2\pi}} 
\, d\psi^1  \, e^S \Bigg/ \int d\theta
\, .
\ee
In the BV formalism the equivalence of the `gauge invariant' form 
of the path integral, where we set all the anti-fields to zero, 
to the Siegel gauge fixed version, 
is usually proved 
at the level of correlation functions\cite{bocc1,bocc2,thorn} for which the overall normalization of the path
integral cancels. Since the normalization is important for us, we have shown 
in appendix \ref{sa} that
the equality of \refb{esigact} and \refb{egaugeinv}
can be understood using the
standard Faddeev-Popov formalism.

We shall now show that \refb{egaugeinv} leads 
to finite unambiguous result. 
Let us first carry out the integral over $\psi^0$ and
$\psi^1$ by taking the integration contours to be the steepest descent contours. 
Both of these
lie along the imaginary axis, and the  final result takes the form:
\be \label{enewN}
\bN = - \zeta\, {1\over \sqrt 2} \, \int d\psi^3 \Bigg/ \int d\theta\, .
\ee
The minus sign in \refb{enewN} is the result of the product of two $i$'s, one from having
to integrate the tachyon $\psi^0$ along the imaginary axis and the other from having to
integrate $\psi^1$ along the imaginary axis. However in the open string field theory, the
reality condition on the mode $\psi^1$ is $(\psi^1)^*=-\psi^1$\cite{9705038}, 
indicating that we should
carry out the path integral over the variable $i\psi^1$ instead of $\psi^1$. 
This would remove
one factor of $i$ from \refb{enewN}. There is however a similar factor of $i$ involved in the
integration over
the gauge 
transformation parameter $\theta$ in \refb{egaugeinv}. These effects cancel each other,
and so we shall proceed with \refb{enewN} without removing any factor of $i$.
This has been discussed in footnote \ref{fo3}.

The mode $\psi^3$ is related by field redefinition to 
the collective mode corresponding to the freedom of translating
the D-instanton along the Euclidean time direction. If $\wt\phi$ denotes the
correctly normalized collective mode that measures the amount of translation along
the time coordinate, then the dependence of any amplitude on 
$\wt \phi$ should be of the form $e^{-i\omega\wt\phi}$ where $\omega$ is the
total energy carried by all the external closed string states. Therefore the relation
between $\psi^3$ and $\wt \phi$ may be found by studying the coupling of $\psi^3$ to
an amplitude and comparing this with the expected coupling of $\wt\phi$ to the same
amplitude. Let us begin with a disk amplitude of a set of closed string states carrying
energies $\omega_1,\omega_2,\cdots$. 
Since the vertex operator of the state associated with $\psi^3$ is given by
$i\, \sqrt 2 \, c \, \p X$, inserting this into this amplitude will correspond to inserting the
integrated vertex operator
\be \label{e26}
i\, \sqrt 2 \,\int \p X(z) dz\, .
\ee
Using the operator product expansion \refb{exxope}, and recalling that when we use the
doubling trick mentioned in footnote \ref{fo1}, insertion of a vertex operator 
$e^{-i\omega_k X(w_k)}$ is implicitly accompanied by its image $e^{i\omega_k X(\bar w_k)}$,
 we get
\ben \label{eopecoll}
&& \left\langle i\, \sqrt 2 \,\int \p X(z) dz\, \prod_k e^{-i\omega_k X(w_k)} \right\rangle
\nonumber\\
&=& i\, \sqrt 2 \,\sum_j \int dz \, \left[
\left\{ {i\omega_j \over 2 (z-w_j)} - {i\omega_j \over 2 (z-\bar w_j)}\right\} 
\, \left\langle\prod_k e^{-i\omega_k X(w_k)} \right\rangle\right]
\nonumber \\
&=& -2\pi i \, {\omega\over \sqrt 2} \, \left\langle \prod_k e^{-i\omega_k X(w_k)}\right\rangle, 
\qquad \omega\equiv\sum_j\omega_j \, .
\een
Since we have not included any dependence on the string coupling in the quadratic terms
in the action, the open string vertex operator \refb{e26} should also carry 
a factor of the open string coupling
$g_o\propto
\sqrt{g_s}$. 
The precise relation between $g_o$ and $g_s$ was determined in \cite{9911116}
and takes the form $g_o^2 = g_s/ (2\pi^2)$ 
in the convention in which the instanton action
is given by $1/g_s$. 
We shall proceed
for now by ignoring the factors of $g_o$ since $g_o$ (in)dependence of $\bN$ 
has already been understood in \cite{private}.
At the end of this
section we shall briefly discuss $g_o$ dependence of different contributions to $\bN$ and
show how they cancel.
\refb{eopecoll} now shows that 
coupling of $\psi^3$ to an amplitude with closed string states carrying total
energy $\omega$ generates a factor of $-\sqrt 2\pi i\omega$. On the other hand, since the 
dependence of an amplitude on the collective coordinate $\wt \phi$ is of the form
$e^{-i\omega\wt\phi}= (1-i\omega\wt\phi+\cdots)$, 
the coupling of $\wt\phi$ to an amplitude with closed string state
carrying
energy $\omega$ generates a factor of $-i\omega$. This gives the identification
of $-\sqrt 2\pi i\omega\psi^3$ to $-i\omega\wt\phi$, in agreement with the results of
\cite{9911116}. Therefore in \refb{enewN} we can
make the replacement:
\be\label{e312}
d\psi^3 = {1\over \sqrt 2 \pi} \, d\wt\phi\, .
\ee
Integration over the collective mode
$\wt\phi$ generates the usual energy conserving delta function $2\pi\delta(\omega)$ 
that is 
part of any amplitude and is not included in the normalization constant
$\bN$.  Therefore we can now express \refb{enewN} as
\be \label{enewerN}
\bN = -\zeta\,  {1\over \sqrt 2} \, {1\over \sqrt 2 \pi} \Bigg/ \int d\theta = -\zeta\, {1\over 2\pi} 
\Bigg/ \int d\theta\, .
\ee

We now turn to the evaluation of $\int d\theta$.
Physically, this gauge transformation is related by field redefinition to the
rigid U(1) gauge transformation that multiplies any state of the 
open string, stretched from the
D-instanton to a second D-instanton, by $e^{i\wt\theta}$. Since $\wt\theta$ 
has period $2\pi$, in order
to determine the range of $\theta$ integral, we need to find the relation between $\theta$
and $\wt\theta$. This in turn can be determined by comparing the string field theory gauge
transformation law generated by $\theta$ to the rigid U(1) gauge transformation with 
parameter $\wt\theta$ for any state of the open string that connects the D-instanton to the
second D-instanton. This is achieved as follows:
\begin{enumerate}
\item As in \cite{2012.11624}, we shall work with a particular mode 
$\xi$ that multiplies the vacuum state
$|0\rangle$ 
of the open string stretched
between the two D-instantons but the relation between $\theta$ and $\wt\theta$ 
is independent of this choice.
The conjugate 
anti-field $\xi^*$ of $\xi$
will multiply the state $c_{-1} c_0 c_1|0\rangle$ of the open string that connects the
second D-instanton to the original D-instanton. 
\item The vertex operators associated with the modes
$\xi$ and $\xi^*$ are accompanied by Chan-Paton 
factors $\pmatrix{0 & 1\cr 0 & 0}$ and $\pmatrix{0 & 0\cr 1 & 0}$ respectively, 
and
the open string mode $\psi^2$, that connects the original D-instanton to itself, carries
Chan-Paton factor  $\pmatrix{1 & 0\cr 0 & 0}$. 
\item It follows from the gauge transformation laws of the string field theory that 
the gauge transformation of $\xi$ under the
gauge transformation generated by $\theta$ 
is given by the second derivative of the  action with
respect to $\xi^*$ and $\psi^2$.
The leading contribution comes from the
$\xi$-$\xi^*$-$\psi^2$ coupling in the action arising from the disk amplitude.
Since two of the three vertex operators -- those associated with $\xi$ and $\psi^2$ are
just identity operators, the coefficient of this term is given by
\be \label{eKapp}
\langle 0| c_{-1}c_0 c_1|0\rangle \, Tr\left[
\pmatrix{0 & 1\cr 0 & 0}
\pmatrix{0 & 0\cr 1 & 0}\pmatrix{1 & 0\cr 0 & 0}
\right] = 1 \, .
\ee
This corresponds to the presence of a term
\be \label{eghost0}
\xi\, \xi^*\, \psi^2
\ee
in the action if we ignore the factors of $g_o$ as before.
\item 
Taking the derivative of \refb{eghost0} with respect to $\xi^*$ and $\psi^2$, we see
that the gauge transformation generated by the parameter
$\theta$ takes the form $\delta\xi = \theta\xi$.
Comparing this with the infinitesimal rigid U(1) transformation 
$\delta\xi=i\wt\theta \xi$. 
we get $\theta=i\wt\theta$.\footnote{This factor of $i$ is the result of imposing wrong
reality condition on the mode $\psi^2$ or equivalently the parameter $\theta$. We have not
corrected it since this cancels the factor of $i$ arising out of the wrong choice of reality
condition for the mode $\psi^1$. This has been discussed below \refb{enewN}. \label{fo3}
} 
\end{enumerate}
This gives
\be \label{efull}
\bN = -\zeta\, {1\over 2\pi} \, \Bigg/ \int d\theta = \zeta\, {i\over 2\pi} \, \Bigg/ \int d\wt\theta 
= \zeta\, {i\over 4\, \pi^2}\, .
\ee

Finally we shall discuss the dependence of $\bN$ on the string coupling. This has already
been fully understood in \cite{private} but we include the discussion here for completeness.
We denote by $g_o= \sqrt{g_s/(2\pi^2)}$ the open string coupling\cite{9911116}.
We shall work in the convention in which the kinetic term of the open string fields
has $g_o$ independent normalization, so that in the Siegel gauge the quadratic part of the
action is $g_o$ independent, in agreement with the $g_o$ independent exponent appearing
in \refb{e8}. In this convention, each open string vertex  operator carries a factor of $g_o$. 
This introduces an additional
factor of $g_o$ in \refb{e26}, \refb{eopecoll} and therefore a factor of $1/g_o$ in 
the right hand sides of \refb{e312}, \refb{enewerN},\refb{efull}. 
On the other hand the disk three point function
of three open string vertex operators now gets a factor of $g_o^{-2}$ from the disk, and 
a factor of $g_o^3$ from the three open string vertex operators, producing a net
factor of $g_o$. Therefore \refb{eghost0} gets a factor of $g_o$, leading to the
gauge transformation law $\delta\xi=g_o \theta\xi$. Therefore we now have 
$g_o\theta=i \, \wt\theta$, leading to an extra factor of $g_o$ on the right hand side
of \refb{efull}. This cancels the earlier factor of $1/g_o$ and leaves the right hand side
of \refb{efull} unchanged. Therefore $\bN$ is $g_o$ independent.

\sectiono{The multiplier} \label{s3.5}

We now turn to the determination of the multiplier $\zeta$.
For this we need to know how the 
steepest descent contour / Lefschetz thimble
passing through the saddle point representing the D-instanton fits inside the
actual integration cycle that computes the full amplitude in string theory\cite{1206.6272,1511.05977,1802.10441}. 
For the case of two dimensional bosonic string theory
this was discussed in \cite{2012.00041} where it was argued that the actual integration contour
contains only half of this thimble. In brief, the argument can be stated as follows.
After integrating out the massive open string modes, the tachyon effective potential
on the D-instanton has a potential $V(\psi^0)$ 
that has a maximum at $\psi^0=0$ describing the
D-instanton and a minimum at some positive value $1/a$ describing the perturbative 
vacuum where the potential vanishes.
The potential is unbounded from below as $\psi^0\to -\infty$. Therefore the integration
contour over $\psi^0$ cannot be taken to be along the real axis all the way to $-\infty$, 
but near the perturbative
vacuum where the potential has a local minimum we expect the contour to lie along
the real axis. If we model the potential as
\be\label{eppp3}
V(\psi^0) = - {1\over 2} (\psi^0)^2 + {1\over 3} a \, (\psi^0)^3 +{1\over 6 \, a^2}\, ,
\ee
then one can easily see that the potential goes to $+\infty$ as we approach the asymptotic 
region within three $60^\circ$ cones, centered  around the lines $\psi^0=r$, $\psi^0=r\, 
e^{2\pi i/3}$ and $\psi^0=r\,  e^{-2\pi i/3}$
for real positive $r$. Therefore we can take the integration contour to interpolate between
the regions $\psi^0=e^{-2\pi i/3}\times\infty$ and $\psi^0=\infty$ as shown in 
Fig.~\ref{figpicardthree} or we can choose
the complex conjugate contour. On the other hand, the
steepest descent contour for the saddle point at $\psi^0=1/a$, representing the perturbative
vacuum, lies along the real $\psi^0$ axis from 0 to $\infty$, while the steepest descent
contour for the saddle point at $\psi^0=0$, representing the D-instanton, consists of a
contour through the origin
that interpolates between the regions around  $\psi^0= e^{-2\pi i/3}\times\infty$ and
$\psi^0=e^{2\pi i/3}\times\infty$. Therefore the integration contour shown in 
Fig.~\ref{figpicardthree}
can be regarded as the union of the steepest descent contour of the saddle point at
$\psi^0=a$, and half of the steepest descent contour of the saddle point at $\psi^0=0$.
This gives
\be \label{ezetafin}
\zeta={1\over 2}\, ,
\ee
and
\be \label{ehalf}
\bN 
= {i\over 8\, \pi^2}\, .
\ee

Let us now comment on the sign of $\bN$ about which we have not been careful so far. 
This clearly depends on the
choice of the full integration contour -- if instead of the contour shown in 
Fig.~\ref{figpicardthree} we choose the complex conjugate contour, the sign of $\bN$ will
change. The actual choice should be dictated by physical
considerations, e.g. if a D-instanton induced amplitude leads to violation of
unitarity, it should be describable by an effective Hamiltonian with negative imaginary
part, reflecting loss of probability due to possible transition to  states that
have not been accounted for in the effective Hamiltonian. 
As discussed in \cite{2012.00041}, 
the choice of sign given in \refb{ehalf} is the correct choice
according to this consideration. Therefore \refb{ehalf} gives the final result for the
normalization constant associated with single D-instanton amplitudes in two dimensional
string theory. This agrees with the result obtained in \cite{1907.07688} 
by comparison with the matrix
model results for the instanton induced amplitudes,
after we multiply this by a factor of $i$ to
compute the normalization of the D-instanton contribution to the S-matrix elements.

\sectiono{Discussion} \label{s4}

The method described here can in principle be applied to other D-instanton systems,
e.g. general $(m,n)$ ZZ-instantons in two dimensional bosonic string theory\cite{1912.07170},
D-instantons in two dimensional type 0B string theory\cite{0307083,0307195} 
and D-instantons
in type IIB string theory\cite{9701093,9712156}. 
Part of the analysis that may be somewhat non-trivial is 
the computation of the multiplier factor $\zeta$, since this requires the knowledge of how the 
steepest descent contour / Lefschetz thimble associated with a particular D-instanton fits
into the full integration contour. For example, in the context of $(m,n)$ ZZ-instantons in two 
dimensional string theory, this will
require the knowledge of 
how the different ZZ-instantons are represented as different extrema in the
configuration space of string fields. However since the multiplier factors are just rational
numbers, we
only need topological information on the locations of various extrema in the 
configuration space of string fields
instead of requiring detailed dynamical information. Therefore we do not consider this to be
an insurmountable problem. Similarly for computing the D-instanton contribution to the
type IIB string theory amplitude, we need to understand how the D-instantons, which in
this case represent complex saddle points, fit inside the integration contour over the string
fields.

\bigskip

\noindent {\bf Acknowledgement:} 
I wish to thank Bruno Balthazar,  
Victor Rodriguez, Xi Yin and Barton Zwiebach for many useful discussions.
This work was
supported in part by the 
J. C. Bose fellowship of 
the Department of Science and Technology, India and the Infosys chair professorship. 

\appendix

\sectiono{Siegel gauge fixing in the Faddeev-Popov formalism} \label{sa}

In our analysis the formal equality 
between the Siegel gauge fixed path integral \refb{esigact} and the
gauge invariant path integral 
\refb{egaugeinv} plays an important role. In this appendix we shall show how this can be
proved using the
standard Faddeev-Popov formalism.

Classical open string field $|\psi_c\rangle$ is an arbitrary state of ghost number 1 
of the open string and
the gauge transformation parameters describing the symmetries of the classical open string
field theory correspond to an arbitrary state $|\theta\rangle$ 
of ghost number $0$. As in \refb{epsir}, we shall work with the restricted string field
carrying $L_0\le 0$. In this case, it is evident from \refb{epsir} that the
gauge transformation
parameter $|\theta\rangle$, carrying ghost number 0, satisfies the condition 
$b_0|\theta\rangle=0$. Let us introduce basis states 
$\{|\phi^{(n)}_r\rangle\}$
of ghost number $n$, satisfying $b_0|\phi^{(n)}_r\rangle=0$. Then the full set of
basis states at ghost number $n$ may be taken to be $\{|\phi^{(n)}_r\rangle\}$ and
$\{c_0|\phi^{(n-1)}_s\rangle\}$. 
We shall further normalize these basis states for $n=0$ and 2 such that
\be
\langle \phi^{(2)}_r| c_0 |\phi^{(0)}_s\rangle =\delta_{rs}\, .
\ee
We can now
expand the restricted classical string field $|\psi_c\rangle$
and the gauge transformation parameter $|\theta\rangle$ as:
\be\label{ea.2}
|\psi_c\rangle =\sum_r \wt
\psi_r |\phi^{(1)}_r\rangle + \sum_s \wh\psi_s c_0 |\phi^{(0)}_s\rangle \, ,
\ee
\be\label{ea.3}
|\theta\rangle = \sum_s \theta_s |\phi^{(0)}_s\rangle\, .
\ee
The gauge invariant classical action up to quadratic order and the linearized 
gauge transformation laws are
given, respectively, by,
\be \label{ea.4}
S_{g.i.}= -{1\over 2} \langle \psi_c|Q_B|\psi_c\rangle\, ,
\ee
and 
\be\label{ea.5}
\delta|\psi_c\rangle = Q_B|\theta\rangle\, .
\ee

We now consider a gauge invariant path integral of the form:
\be\label{ea.6}
I_{g.i.} = \int \prod_r d\wt\psi_r \prod_s d\wh\psi_s \, e^{S_{g.i.}} \Bigg/ \int \prod_s d\theta_s
\ee
We can evaluate this by gauge fixing in the Siegel gauge, by setting $\wh \psi_s=0$.
This introduces a factor of $\prod_s \delta(\wh\psi_s)$ in the path integral, accompanied by the
appropriate determinant. To find the determinant we need to examine the gauge
transformation law of $\wh\psi_s$. Substituting \refb{ea.2}, \refb{ea.3} 
into \refb{ea.5} and taking the
inner product of the resulting equation with $\langle\phi^{(2)}_s|$, we get
\be\label{ea.7}
\delta \wh\psi_s = \langle\phi^{(2)}_s| Q_B |\theta\rangle = M_{su} \theta_u, \qquad
M_{su} =  \langle\phi^{(2)}_s| Q_B | \phi^{(0)}_u\rangle\, .
\ee
The determinant entering the integrand of the path integral is $\det M$. We can 
express this as a path integral over a pair of grassmann odd ghost variables 
$\{p_s\},\{q_s\}$. This gives the gauge fixed path integral:
\be \label{ea.9}
I_{g.f.} = \int \prod_r d\psi_r  \prod_s dp_s d q_s \, e^{S_{g.i.} 
+ S_{ghost}}\bigg|_{\wh\psi_s=0},
\qquad S_{ghost} = -\sum_{s,u} p_s M_{su} q_u =-\langle P| Q_B |Q\rangle \, ,
\ee
where we have introduced 
\be
|P\rangle = \sum_s p_s |\phi^{(2)}_s\rangle, \qquad |Q\rangle = \sum_s 
q_s |\phi^{(0)}_s\rangle\, .
\ee

The equality of $I_{g.i.}$ and $I_{g.f.}$ give in 
\refb{ea.6} and \refb{ea.9} in the context of two dimensional string theory
gives the equality of \refb{egaugeinv} and \refb{esigact}. To see this note that in this case
the master field $|\psi\rangle$ 
contains fields of all ghost number without any gauge condition, and the
master action is $-{1\over 2}\langle\psi|Q_B|\psi\rangle$. When we pick the Lagrangian submanifold
in which we set the modes of ghost number $\ge 2$ to 0, the master action reduces to
\refb{ea.4} and \refb{egaugeinv} may be interpreted as $\zeta/(2\pi)$ times $I_{g.i.}$
given in \refb{ea.6}.
On the other hand
when we pick the Lagrangian submanifold by imposing the Siegel gauge condition, 
the master action reduces to $S_{g.i.}+S_{ghost}$ given in \refb{ea.9} and 
\refb{esigact} may be interpreted as $\zeta/(2\pi)$ times $I_{g.f.}$ given in
\refb{ea.9}.

The equality of $I_{g.i.}$ and $I_{g.f.}$ that we have established is formal,
since in the context in which we apply this, 
there are component fields that multiply basis states of vanishing $L_0$ eigenvalue.
This means that the Siegel gauge 
action becomes independent of those fields, making the path
integral \refb{ea.9} ill-defined.
This can be traced to the fact that the Siegel gauge $\wh\psi_r=0$ 
is a singular gauge choice due to
the appearance of zero eigenvalues of the matrix $M$ introduced in \refb{ea.7}.
However the original gauge invariant path integral
\refb{ea.6} is well defined. The point of view we take is that \refb{ea.6} is the
proper definition of the path integral, and the problem we face with \refb{ea.9} arising 
in the world-sheet formalism is due to illegal choice of gauge.\footnote{As discussed below
\refb{e8}, $I_{g.f.}$ can be made well-defined by working with open strings stretched
between a pair of separated D-instantons. In that case the equality between $I_{g.i.}$ 
and $I_{g.f.}$ becomes an identity.
}

The analysis described above also resolves an apparent puzzle that arises out
of the equality of \refb{esigact} and \refb{egaugeinv}.  Since the
integral in \refb{esigact} has equal number of bosonic and fermionic variables, it
remains unchanged if we multiply $S$ in
the exponent by some constant $C$. Following the
logic that led from \refb{esigact} to \refb{egaugeinv}, we can see that the effect of this
rescaling is to multiply the exponent in \refb{egaugeinv} by the same constant $C$.
Since in this expression both $\psi^0$ and $\psi^1$ represent bosonic variables, this
will produce a factor of $1/C$ on the right hand side of \refb{enewN}
that carries over to the right hand sides of 
\refb{enewerN} and \refb{efull}. So the question is: what compensates this factor?

The answer to this question comes from \refb{ea.7} and \refb{ea.9}. If the 
exponent of \refb{esigact} is multiplied by $C$, then the ghost action involving the
modes $p_s$, $q_s$, which correspond to $\psi^2$ and $\psi_1$ in \refb{esigact},
also gets multiplied by $C$. Therefore the matrix $M$ in \refb{ea.7} must be
multiplied by $C$. This can happen if we include an extra factor of $C$ in the gauge 
transformation law \refb{ea.5}. If we denote by $\theta'$ the new gauge transformation
parameter, then it is related to the old gauge transformation parameter $\theta$ by
$\theta=C\theta'$. Therefore the $\int d\theta'$ factor that now appears in the denominator
can be identified to $\int d\theta/C$, and we get an extra multiplicative factor of $C$ 
on the right hand side of \refb{enewN}. This cancels the extra factor of $1/C$
coming from the integration over $\psi^0$ and $\psi^1$.


\begin{thebibliography}{99}

\bibitem{9407031} 
  J.~Polchinski,
  ``Combinatorics of boundaries in string theory,''
  Phys.\ Rev.\ D {\bf 50}, R6041 (1994)
  doi:10.1103/PhysRevD.50.R6041
  [hep-th/9407031].
  %%CITATION = doi:10.1103/PhysRevD.50.R6041;%%

\bibitem{9701093} 
  M.~B.~Green and M.~Gutperle,
  ``Effects of D instantons,''
  Nucl.\ Phys.\ B {\bf 498}, 195 (1997)
  doi:10.1016/S0550-3213(97)00269-1
  [hep-th/9701093].
  %%CITATION = doi:10.1016/S0550-3213(97)00269-1;%%


\bibitem{1907.07688} 
  B.~Balthazar, V.~A.~Rodriguez and X.~Yin,
  ``ZZ Instantons and the Non-Perturbative Dual of c = 1 String Theory,''
  arXiv:1907.07688 [hep-th].
  %%CITATION = ARXIV:1907.07688;%%

\bibitem{wittensft} 
  E.~Witten,
  ``Noncommutative Geometry and String Field Theory,''
  Nucl.\ Phys.\ B {\bf 268}, 253 (1986).
  %%CITATION = NUPHA,B268,253;%%

\bibitem{9206084} 
  B.~Zwiebach,
  ``Closed string field theory: Quantum action and the B-V master equation,''
  Nucl.\ Phys.\ B {\bf 390}, 33 (1993)
  doi:10.1016/0550-3213(93)90388-6
  [hep-th/9206084].
  %%CITATION = doi:10.1016/0550-3213(93)90388-6;%%

\bibitem{9705241} 
  B.~Zwiebach,
  ``Oriented open - closed string theory revisited,''
  Annals Phys.\  {\bf 267}, 193 (1998)
  doi:10.1006/aphy.1998.5803
  [hep-th/9705241].
  %%CITATION = doi:10.1006/aphy.1998.5803;%%



\bibitem{1703.06410} 
  C.~de Lacroix, H.~Erbin, S.~P.~Kashyap, A.~Sen and M.~Verma,
  ``Closed Superstring Field Theory and its Applications,''
  Int.\ J.\ Mod.\ Phys.\ A {\bf 32}, no. 28n29, 1730021 (2017)
  doi:10.1142/S0217751X17300216
  [arXiv:1703.06410 [hep-th]].
  %%CITATION = doi:10.1142/S0217751X17300216;%%

\bibitem{1905.06785}
T.~Erler,
``Four Lectures on Closed String Field Theory,''
Phys. Rept. \textbf{851}, 1-36 (2020)
doi:10.1016/j.physrep.2020.01.003
[arXiv:1905.06785 [hep-th]].

\bibitem{okawa}
Y.~Okawa,
``Complete formulation of superstring field theory,''
doi:10.1142/9789813144613\_0002

 \bibitem{1908.02782} 
  A.~Sen,
  ``Fixing an Ambiguity in Two Dimensional String Theory Using String Field Theory,''
  JHEP {\bf 2003}, 005 (2020)
  doi:10.1007/JHEP03(2020)005
  [arXiv:1908.02782 [hep-th]].
  %%CITATION = doi:10.1007/JHEP03(2020)005;%%

\bibitem{2002.04043} 
  A.~Sen,
  ``D-instanton Perturbation Theory,''
  JHEP \textbf{08}, 075 (2020)
doi:10.1007/JHEP08(2020)075
[arXiv:2002.04043 [hep-th]].
%%CITATION =doi:10.1007/JHEP08(2020)075;%%

\bibitem{2003.12076}
A.~Sen,
``Divergent to Complex Amplitudes in Two Dimensional String Theory,''
[arXiv:2003.12076 [hep-th]].

\bibitem{2012.11624}
A.~Sen,
``D-instantons, String Field Theory and Two Dimensional String Theory,''
[arXiv:2012.11624 [hep-th]].
 
\bibitem{9111035}
G.~W.~Moore, M.~Plesser and S.~Ramgoolam,
``Exact S matrix for 2-D string theory,''
Nucl. Phys. B \textbf{377}, 143-190 (1992)
doi:10.1016/0550-3213(92)90020-C
[arXiv:hep-th/9111035 [hep-th]].

\bibitem{bv1} 
  I.~A.~Batalin and G.~A.~Vilkovisky,
  ``Gauge Algebra and Quantization,''
  Phys.\ Lett.\  {\bf 102B}, 27 (1981).
  doi:10.1016/0370-2693(81)90205-7
  %%CITATION = doi:10.1016/0370-2693(81)90205-7;%%

\bibitem{bv2} 
  I.~A.~Batalin and G.~A.~Vilkovisky,
  ``Quantization of Gauge Theories with Linearly Dependent Generators,''
  Phys.\ Rev.\ D {\bf 28}, 2567 (1983)
  Erratum: [Phys.\ Rev.\ D {\bf 30}, 508 (1984)].
  doi:10.1103/PhysRevD.28.2567, 10.1103/PhysRevD.30.508
  %%CITATION = doi:10.1103/PhysRevD.28.2567, 10.1103/PhysRevD.30.508;%%

\bibitem{henn}
M.~Henneaux and C.~Teitelboim, ``Quantization of Gauge Systems'', Princeton
University Press, Princeton, New Jersey, 1992.


\bibitem{bocc1}
M.~Bochicchio,
``Gauge Fixing for the Field Theory of the Bosonic String,''
Phys. Lett. B \textbf{193}, 31-36 (1987)
doi:10.1016/0370-2693(87)90451-5

\bibitem{bocc2}
M.~Bochicchio,
``String Field Theory in the Siegel Gauge,''
Phys. Lett. B \textbf{188}, 330 (1987)
doi:10.1016/0370-2693(87)91391-8

\bibitem{thorn}
C.~B.~Thorn,
``STRING FIELD THEORY,''
Phys. Rept. \textbf{175}, 1-101 (1989)
doi:10.1016/0370-1573(89)90015-X

\bibitem{9205088}
A.~S.~Schwarz,
``Geometry of Batalin-Vilkovisky quantization,''
Commun. Math. Phys. \textbf{155}, 249-260 (1993)
doi:10.1007/BF02097392
[arXiv:hep-th/9205088 [hep-th]].


\bibitem{1206.6272}
M.~Marino,
``Lectures on non-perturbative effects in large $N$ gauge theories, matrix models and strings,''
Fortsch. Phys. \textbf{62}, 455-540 (2014)
doi:10.1002/prop.201400005
[arXiv:1206.6272 [hep-th]].

\bibitem{1511.05977}
G.~V.~Dunne and M.~Unsal,
``What is QFT? Resurgent trans-series, Lefschetz thimbles, and new exact saddles,''
PoS \textbf{LATTICE2015}, 010 (2016)
doi:10.22323/1.251.0010
[arXiv:1511.05977 [hep-lat]].


\bibitem{1802.10441}
I.~Aniceto, G.~Basar and R.~Schiappa,
``A Primer on Resurgent Transseries and Their Asymptotics,''
Phys. Rept. \textbf{809}, 1-135 (2019)
doi:10.1016/j.physrep.2019.02.003
[arXiv:1802.10441 [hep-th]].

\bibitem{9911116}
A.~Sen,
``Universality of the tachyon potential,''
JHEP \textbf{12}, 027 (1999)
doi:10.1088/1126-6708/1999/12/027
[arXiv:hep-th/9911116 [hep-th]].

\bibitem{0810.1737}
M.~Kiermaier, Y.~Okawa and B.~Zwiebach,
``The boundary state from open string fields,''
[arXiv:0810.1737 [hep-th]].


\bibitem{2012.00041}
A.~Sen,
``Cutkosky Rules and Unitarity (Violation) in D-instanton Amplitudes,''
[arXiv:2012.00041 [hep-th]].

\bibitem{1912.07170} 
  B.~Balthazar, V.~A.~Rodriguez and X.~Yin,
  ``Multi-Instanton Calculus in $c = 1$ String Theory,''
  arXiv:1912.07170 [hep-th].
  %%CITATION = ARXIV:1912.07170;%%

\bibitem{1912.05463} 
  H.~Erbin, C.~Maccaferri and J.~Vosmera,
  ``Localization of Effective Actions in Heterotic String Field Theory,''
  arXiv:1912.05463 [hep-th].
  %%CITATION = ARXIV:1912.05463;%%

\bibitem{2006.16270}
H.~Erbin, C.~Maccaferri, M.~Schnabl and J.~Vosmera,
``Classical algebraic structures in string theory effective actions,''
[arXiv:2006.16270 [hep-th]].

\bibitem{9705038}
M.~R.~Gaberdiel and B.~Zwiebach,
``Tensor constructions of open string theories. 1: Foundations,''
Nucl. Phys. B \textbf{505}, 569-624 (1997)
doi:10.1016/S0550-3213(97)00580-4
[arXiv:hep-th/9705038 [hep-th]].

\bibitem{private}
B.~Balthazar, V.~A.~Rodriguez and X.~Yin, private communication.

\bibitem{0307083}
T.~Takayanagi and N.~Toumbas,
``A Matrix model dual of type 0B string theory in two-dimensions,''
JHEP \textbf{07}, 064 (2003)
doi:10.1088/1126-6708/2003/07/064
[arXiv:hep-th/0307083 [hep-th]].

\bibitem{0307195}
M.~R.~Douglas, I.~R.~Klebanov, D.~Kutasov, J.~M.~Maldacena, E.~J.~Martinec and N.~Seiberg,
``A New hat for the c=1 matrix model,''
[arXiv:hep-th/0307195 [hep-th]].

\bibitem{9712156}
M.~Gutperle,
``Aspects of D instantons,''
NATO Sci. Ser. C \textbf{520}, 411-422 (1999)
[arXiv:hep-th/9712156 [hep-th]].


\end{thebibliography}
\end{document}